\documentclass[runningheads]{llncs}
\usepackage{graphicx}
\usepackage{comment}
\usepackage{amsmath,amssymb} 
\usepackage{color}

\usepackage[width=122mm,left=12mm,paperwidth=146mm,height=193mm,top=12mm,paperheight=217mm]{geometry}

\begin{document}

\pagestyle{headings}
\def\ECCVSubNumber{5202}  
\title{STD-Net: Structure-preserving and Topology-adaptive Deformation Network for 3D Reconstruction from a Single Image} 
\makeatletter
\def\blfootnote{\xdef\@thefnmark{}\@footnotetext}
\makeatother

\titlerunning{ECCV-20 submission ID \ECCVSubNumber} 
\authorrunning{ECCV-20 submission ID \ECCVSubNumber} 
\author{Anonymous ECCV submission}
\institute{Paper ID \ECCVSubNumber}

\titlerunning{STD-Net}
%


\author{Aihua Mao$^{1\,\star}$, ~~Canglan Dai$^{1\,\star}$, ~~Lin Gao$^{2\,\dag}$, \\Ying He$^{3}$, ~~Yong-jin Liu$^{4}$}
\authorrunning{A.H. Mao et al.}
\institute{\textsuperscript{1}School of Computer Science and Engineering,\\ South China University of Technology\\ 
\textsuperscript{2}Chinese Academy of Sciences (CAS)\\
~\textsuperscript{3}Nanyang Technological University 
~\textsuperscript{4}Tsinghua University\\
\email{ahmao@scut.edu.cn}
~~~\email{dcl\_1016@outlook.com}
~~~\email{gaolin@ict.ac.cn}\\
\email{yhe@ntu.edu.sg}
~~~\email{liuyongjin@tsinghua.edu.cn}
}

\blfootnote{$^\star$ indicates equal contributions.}
\blfootnote{$^\dag$ indicates corresponding author.}


\maketitle

\begin{abstract}
   3D reconstruction from a single view image is a long-standing problem in computer vision. Various methods based on different shape representations (such as point cloud or volumetric representations) have been proposed. However, the 3D shape reconstruction with fine details and complex structures are still challenging and have not yet be solved. Thanks to the recent advance of the deep shape representations, 
   it becomes promising to learn the structure and detail representation using deep neural networks. 
    In this paper, we propose a novel method called STD-Net to reconstruct the 3D models utilizing the mesh representation that is well suitable for characterizing complex structure and geometry details. 
    To reconstruct complex 3D mesh models with fine details, our method consists of (1) an auto-encoder network for recovering the structure of an object with bounding box representation 
    from a single image, (2) a topology-adaptive graph CNN for updating vertex position for meshes of complex topology, and (3) an unified mesh deformation block that deforms the structural boxes into structure-aware meshed models. Experimental results on the images from ShapeNet show that our proposed STD-Net has better performance than other state-of-the-art methods on reconstructing 3D objects with complex structures and fine geometric details. 
\keywords{We would like to encourage you to list your keywords within
the abstract section}
\end{abstract}

\section{Introduction}

3D reconstruction plays an essential role in various tasks in computer vision and computer graphics. Traditional approaches mainly utilize the stereo correspondence based on multi-view geometry, but are restricted to the coverage provided by the input views. Such limitation causes the single-view reconstruction particularly tricky due to the lack of correspondence with other views and large occlusions. With the availability of large-scale 3D shape database \cite{shapenet2015}, shape information can be efficiently encoded in a deep neural network, enabling faithful 3D reconstruction even from a single image. Although many 3D representations (such as voxel-based and point cloud representations), have been utilized for 3D reconstruction, they are not efficient to express the surface details of the shape and may generate part-missing or broken structures due to the high computational cost and memory storage. On the contrary, the triangle mesh is well known by its high efficiency for modelling geometric details, it has attracted considerable attention in computer vision and computer graphics.

Recently, the mesh-based 3D methods have been explored with the deep learning technology\cite{Wang2018:Pixel2Mesh,Groueix2018:AtlasNet}. The triangle mesh can be represented by the graph-based neural network~\cite{tan2018mesh,Litany18}. Although these methods can reconstruct the surface of the object, the reconstruction results are still limited to some categories of 3D models and miss structural information of the object. 
In literature, structure recovery of 3D shapes is mostly studied with non-deep learning approaches, which is due to the lack of a structural representation of 3D shape suitable for deep neural networks. Thus it is necessary to build up a deep neural network that can directly recover the 3D shape structure of an object from a single RGB image. Recent works show that cuboid primitives can be an good choice for structure representation \cite{mo2019structurenet,Li2017:GRASS}. Meanwhile, the cuboid structure representation can recover the complete models with part relationship, while the results estimated by 3D-GAN \cite{3dgan} may lose some parts. However, it still lacks of the surface details of the object. In order to delicately express the shape's structural information through advancing the cuboid representation, we propose a deep neural network to reconstruct 3D objects in the mesh representation level, which makes it feasible to express the complex structure and fine-grained surface details of the 3D objects simultaneously. 

So far mesh-based deep learning approaches mostly rely on the Graph Convolution Network (GCN) \cite{Wang2018:Pixel2Mesh,Groueix2018:AtlasNet,smith19a:GEOMetrics}. The common practice of these methods is to use GCN to deform a pre-defined mesh which is generally a sphere, or deform a set of primitives (square planar) to form 3D shapes. Although GCN is effective in many classification and regression tasks, it may be inadequate for reconstruction, generation or 3D structure analysis, due to that it may cause over-smoothing when aggregating neighbouring information at the vertex level. More importantly, the existing GCN-based methods always deal with fixed-topology meshes, while the cuboid structure are naturally suitable for representing variable topological mesh, which causes that the existing GCNs are not suitable for the cuboid representation. In this paper we aim to address these challenging issues and reconstruct 3D meshed models with adaptive topology and fine-grained geometric details from a single RGB image. The key idea for the single view reconstruction is to obtain the structural representation of 3D objects with cuboids and then deform them into concrete meshes through an integrated deep neural network (called STD-Net), by which a 3D structure-preserving object model can be reconstructed from a single RGB image. The contributions of this paper are summarized as:

\begin{figure*}
   \begin{center}
   \includegraphics[width=\linewidth]{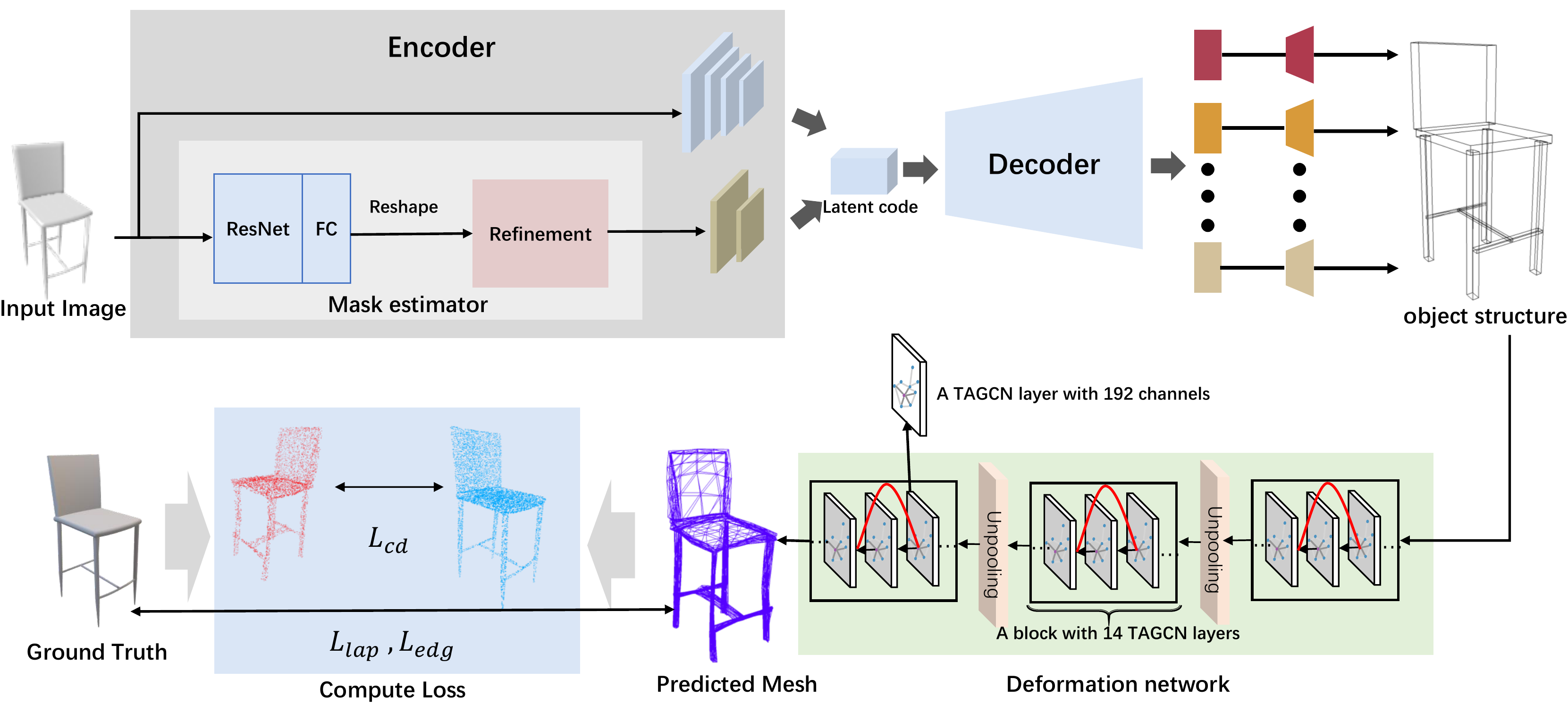}
   \end{center}
      \caption{The overview of STD-Net.Given an input image, firstly an auto-encoder architecture is employed to obtain the bounding boxes covering the parts of the object. Then, multiple mesh deformation and unpooling operations are implemented to progressively deform the mesh vertices
      and update the topology to approximate the target 3D object surface.}
   \label{fig:short}
\end{figure*}

\begin{itemize}
  \item We propose an integrated learning framework that takes an single view-free image as input and construct a 3D meshed models with complex structure and fine-grained geometric details. 
  \item We represent the 3D object structure by recovering the cuboids bounding boxes that can delicately express the rich structure information.  
  \item We construct a topology-adaptive mesh deformation network which releases the limitation with the fixed-topology of input graph and is able to reconstruct shapes with various topologies.     
\end{itemize}

\section{Related Work}

\subsection{3D Object Reconstruction}

The voxel representation has been widely used for 3D shape generation and understanding with the neural network \cite{Tatarchenko2017,Tulsiani2017,Choy2016:3D-R2N2,Huang2018,Han2017}. 
Although it is simple for implementation and has a good compatibility, voxel representation are limited by the resolution and computation cost of 3D convolution. Recently, octree is exploited to develop computationally efficient approaches \cite{Riegler2017:octree,Tatarchenko2017,Hane2018}. 

Another popular representation for 3D shapes is point cloud \cite{Qi2017,Achlioptas2018,Yang2018}, 
which describes 3D shapes through a set of points in the 3D space. Naturally, the point cloud can represent the surface information of 3D objects with no local connections between points, which makes it flexible with sufficient degrees of freedom to match the 3D shape with arbitrary topology. However, the irregular structure of point clouds also leads to a great challenge for deep learning, which is only able to produce relatively coarse geometry and can not be used to directly recover a detailed 3D shape. 

Recently, the mesh models have been introduced in deep learning for 3D generation and reconstruction tasks. Pontes et al. \cite{pontes2017image2mesh} proposed a learning framework to infer 3D mesh models from a single image using a compact mesh representation. Wang et al. \cite{Wang2018:Pixel2Mesh} generated 3D mesh models from a RGB image by deforming an initial ellipsoid mesh from coarse to fine. Groueix et al. \cite{Groueix2018:AtlasNet} deformed over a set of primitives to generate parametric surface elements for 3D shapes. Smith et al. \cite{smith19a:GEOMetrics} presented a approach for adaptive mesh reconstruction, which focuses on exploiting the geometric structure of graph encoded objects. Some methods attempt to reconstruct category-specific mesh which is parameterized by a learned mean shape \cite{Kanazawa_2018_ECCV:category-specific,groueix2018b:3D-CODED}. Our method is mostly related to the recent works \cite{Wang2018:Pixel2Mesh,smith19a:GEOMetrics}, which deform a generic pre-defined mesh to form 3D surfaces. However, the work \cite{Wang2018:Pixel2Mesh} did not concern the  structure information in the 3D representation, hence it can not reconstruct the structure-welled objects. The work \cite{smith19a:GEOMetrics} takes additional voxels to provide the global structure and can not express the local structure. Moreover, both of them require a fixed-topology mesh model, while our method allows the input meshes with different topological types, which is able to generate more delicate structure for the 3D shape.


\subsection{3D Structure Learning}
There are already many works based on learning methods which generate the 3D models by the manners of voxels, point cloud and meshes. However, the outputs of them are the non-structured models. The main reason is that it sill lack of an effective structural representation for 3D shapes currently. Some works attempt to address this issue through non-deep-learning approaches. For example, Xu et al. \cite{Xu2011} modeled 3D objects from photos by utilizing an available set of 3D candidate models. 
Huang et al. \cite{Huang2015} jointly analyzed a collection of images and together reconstructed the 3D shapes from existing 3D models collections. 

Recently, researchers have explored the possibility in expressing the 3D structure through the learning methods. 
\cite{Tulsiani2017:abstractions} proposed a deep architecture to map a 3D volume to a 3D cuboid primitive, which provides potentials to automatically discover and exploit consistent structure in the data. Niu et al. \cite{Niu2018:Im2struct} developed a convolutional-recursive auto-encoder comprised of structure parsing of a 2D image and structure recovering of a cuboid hierarchy. Gao et al. {\cite{Gao2019SDMNETDG:SDM}} reported a deep generative neural network to produce structured deformable meshes. Our method adopts the ideas from these works and proposes a cost-effective auto-encoder to generate cuboid bounding boxes expressing the hierarchy structure of the 3D objects. The cuboid bounding boxes are further meshed and input into the learning network for 3D model generation. 

\subsection{Graph Convolution Neural Network}
Graph convolution neural network \cite{Scarselli2009} becomes a popular choice in 3D shape analysis \cite{Yi2017} and 3D reconstruction \cite{Wang2018:Pixel2Mesh}. Unlike traditional CNN which is defined on regular grids (e.g., 2D pixels and 3D voxels), GCN regards the mesh as a graph and learns features using graph convolution. 
The potential of graph convolution is to capture the structural relations among node of data. But, the irregular structure attributes of graphs leads the huge challenges to the convolutions on graphs. 
The issues in deep learning on graphs are mainly from the complex and diverse connectivity patterns in the graph-structured data. 

Defferrard et al. \cite{Defferrard2016} proposed a approach to approximate the filters by applying the Chebyshev polynomials applied on the Laplacian operator.
This approximation was further simplified by \cite{Kipf2017}, which propose a filter in the spatial domain and achieves good performance on classification task. 
The assumption of symmetric adjacency matrix in above two spectral based methods restricts the application to undirected graph-structured data. The issue of extending CNNs from grid-structured data to arbitrary graph-structured data remains unsolved. 
For adapting to the various topologies of the graph data, Du et al. \cite{Du2018TopologyAG:TAGCN} implement the graph convolution by a set of fixed-size learnable filters.
We adopts the idea of TAGCN in \cite{Du2018TopologyAG:TAGCN} to construct GCN to achieve deformation for meshed cuboid bounding boxes with different structure topology types.

\section{Methodology}
\subsection{Network Architecture}\label{overall}
The network architecture of our method (STD-Net) is shown in Fig. \ref{fig:short}, which is composed of two parts: \textit{structure recovery network}  
and \textit{mesh deformation network}. The structure recovery network is designed with an auto-encoder to predict the 3D structure of the object from a single RGB image. It can generate the object's hierarchy cuboid bounding boxes, which delicately describe the structure information in detail. These bounding boxes are further meshed and put them into the GCNs in the next phase. 

The mesh deformation network aims to deform the input bounding box into a structure-preserving shape. It consists of three blocks intersected by two graph unpooling layers. Each block has the same network structure containing 14 TAGCN layers with 192 channels that accept variable topologies of the graph. The unpooling layer works to add the number of vertices to handle fine-grained geometric details. The detailed explanation for two network parts are discussed in the following Sections.


\subsection{Structure Recovery Network}\label{sec:srn}
Recently, shape abstraction \cite{mo2019structurenet,Niu2018:Im2struct} has been used to discover the high-level structure in un-annotated point clouds and images. These works inspire us to use the decoder in shape abstraction to recover the structure of the object. In our method, an encoder is first employed to map a shape (represented as a hierarchy of $n-$ array graphs or cuboids) to a latent feature vector $z$. Then, a decoder transforms the feature vector $z$ back into a shape (also represented as a hierarchy of graphs or cuboids). The structure of a object is represented by a hierarchy graph, and every node is represented with a bounding box.

For the encoder part, the structure information of image goes through a CNN network and is transferred into a latent code as features. 
For the decoder part, an recursively network unfolds the features into a hierarchical organization of the bounding boxes, which are the recovered structure.

\subsubsection{Encoder}
The encoder takes a 2D RGB image as input to obtain the object structure. It contains two parts. The first part is the contour estimator to estimate the contour of the object that provides strong cues for understanding shape structures in the image. Inspired by the multi-scale network for detailed depth estimation \cite{Li2017}, we design a two-scale CNNs, where the first one captures the information of the whole image and the second one produces a detailed mask map with a quarter of the input resolution.
The second part is to fuse the features of the input image, which is conducted by two convolution channels. One channel takes the binary mask as input, followed by two convolution layers. The other channel takes the CNN feature of the original image (extracted by a ResNet18) as input. The output feature from the two channels are then concatenated, and further encoded into a $n-$D code after two fully connected layers, capturing the object structure information from the input image.
\subsubsection{Decoder}We adopt a recursive network architecture \cite{mo2019structurenet} as a hierarchy of graphs in the decoder, which is more cost-effective than GRASS \cite{Li2017:GRASS}.
In the whole structure recovery network, the latent code can be regarded as the root features of the structure tree. The decoder gradually decodes the node feature code into a hierarchy of features until reaching the leaf nodes where each of them can
be parsed into a vector of box parameters. In more detail, there are two types of decoders in the whole decode pipeline. The bounding box decoder is implemented as a multi-layer perception (MLP) with two layers, which transforms a feature vector to a bounding box. The graph decoder transforms a parent feature vector back into the child graph. 
\begin{figure*}
  \begin{center}
   \includegraphics[width=\linewidth]{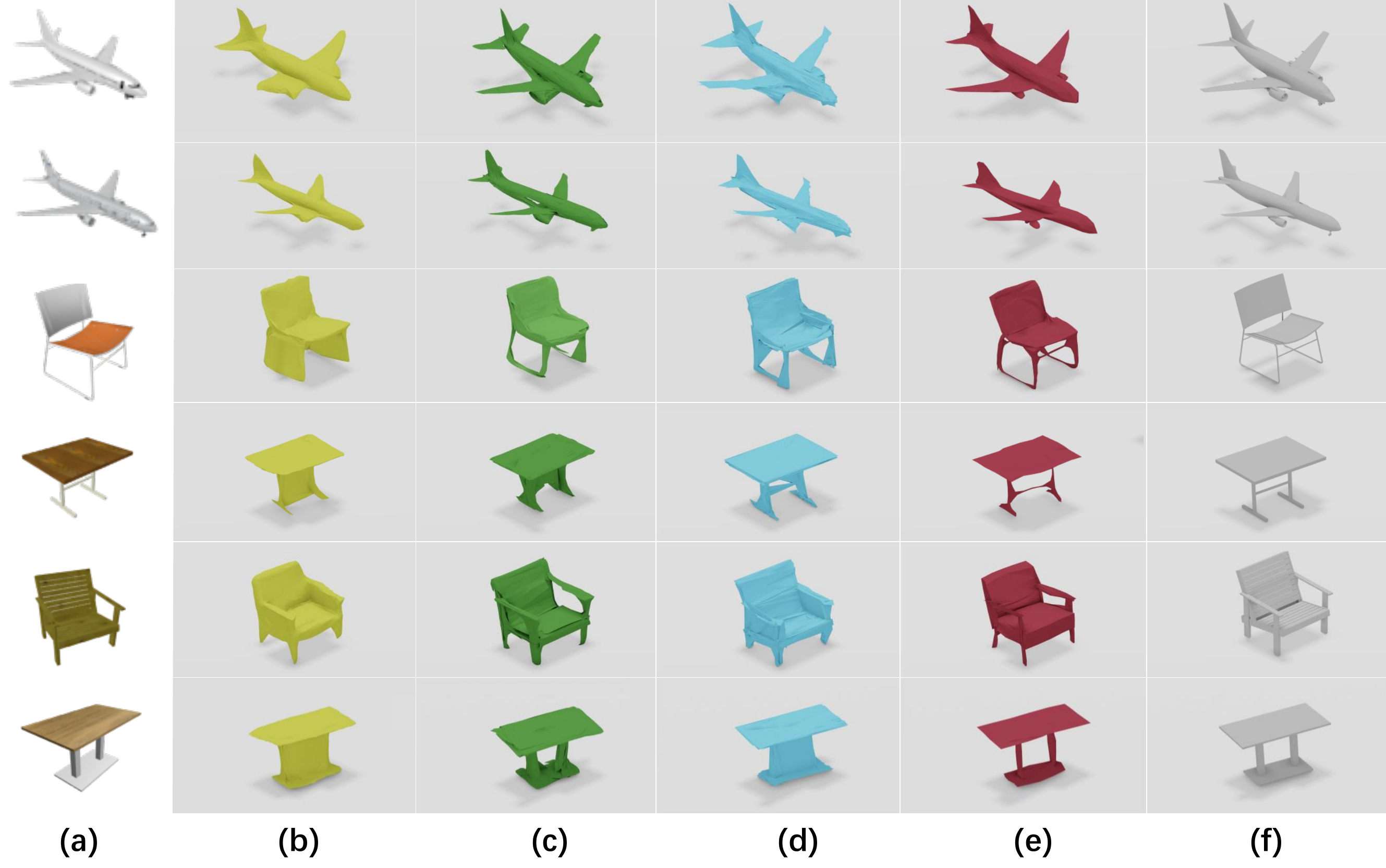}
  \end{center}
     \caption{Reconstruction results in three categories: Airplane, Table and Chair. (a) Input image; (b) Pixel2Mesh; (c) AtlasNet-25; (d) GEOMetrics; (e) Ours; (f) Ground Truth.}
  \label{fig:result}
\end{figure*}

\subsection{Mesh Deformation Network}\label{sec:3.3}
Given a bounding box $B$ generated from Section \ref{sec:srn} and the ground truth $S$, our goal is to deform the bounding box to make it to be as close to the ground truth shape as possible. Figure. \ref{fig:short} depicts our mesh deformation module, which takes a meshed bounding box defined by a set of vertex positions as input and output predicted deformed mesh. 
In the mesh deformation network, three deformation blocks are intersected by two graph unpooling layers to progressively achieve the mesh deformation. Each deformation block takes the graph (representing the mesh with the 3D shape feature attached on vertices) as input. The unpooling layer is to increase the number of vertices, which can increase the capacity of handling fine-grained geometric details. 


The three deformation blocks having same architecture containing 14 TAGCN layers with 192 channels. The first block takes the initial meshed bounding box as input, and the remaining two blocks take the output from the previous unpooling layer as input. Inspired by the TAGCN \cite{Du2018TopologyAG:TAGCN}, we only care about a small local region around the vertex (see Fig \ref{fig:tagcn}). By defining a \textit{path} of length $m$ on a graph as a sequence $v = (v_0 , v_1 , ..., v_m)$, $v_k \in  V$, the small region can be determined by the node \textit{path}. The convolution formula used in each convolution layer is 

\begin{align}\label{eqn:conv}
   X_{l+1} = f(A^KX_lW_K^l + \dots + A^1X_lW_1^l + X_lW_0^l)
\end{align}%
where $A$ is the adjacency matrix , $X_l$ is the input vector in $l^{th}$ convolution layer, $f$ is the nonlinear activation function , and $W_k$ are the learnable weights. Through experiments, it has been shown that $k = 2$ can achieve good performance.
The convolution defined in Eq.(\ref{eqn:conv}) is similar to the traditional convolution operations. In the convolution layer of traditional CNN, the receptive filed is a small rectangle in the grid. In GCN, the receptive field is also a small region around the vertex. The operation in Eq.(\ref{eqn:conv}) calculates a linear combination of the signals of nodes $k$ hops away from the start node. 
Moreover, we propose a deep network with several shortcut connections, which can alleviate the over-smoothing in GCN. In addition to the output of the network, there is a branch which applies an extra TAGCN layer to the last layer and outputs the 3D coordinates. Similar to Pixel2Mesh, we add a new vertex at the center of the each edge to form each unpooling layer, which can increase the number of vertices in GCN.

\begin{figure}[]
  \begin{center}
     \includegraphics[width=1\linewidth]{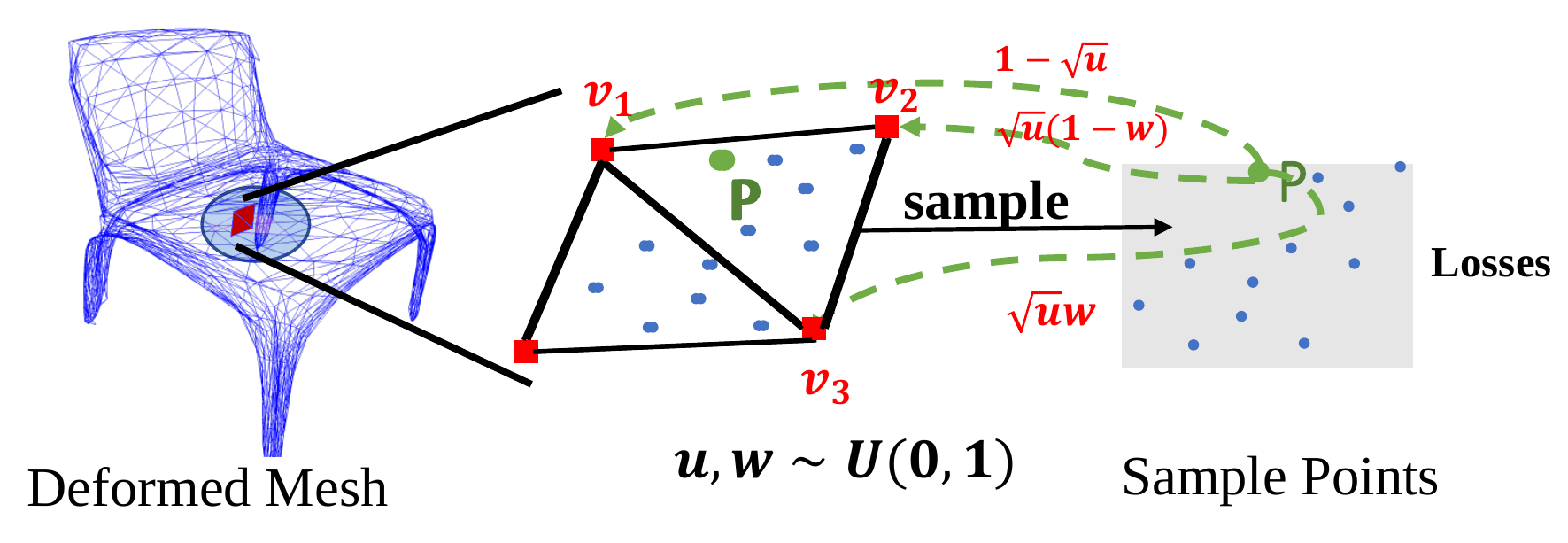}
  \end{center}
     \caption{Sampling strategy from predicted mesh.}
  \label{fig:sample}
  \end{figure}

\subsubsection{Losses} The mesh deformation network is supervised by a hybrid losses including the chamfer distance loss, Laplacian loss and edge length loss. The chamfer distance (CD) measures the nearest neighbor distances between two point sets. 
We minimize the two directional distances between the deformed bounding boxes and the groundtruth shape. The chamfer distance loss is defined as:

\begin{figure}
  \begin{center}
   \includegraphics[width=.7\linewidth]{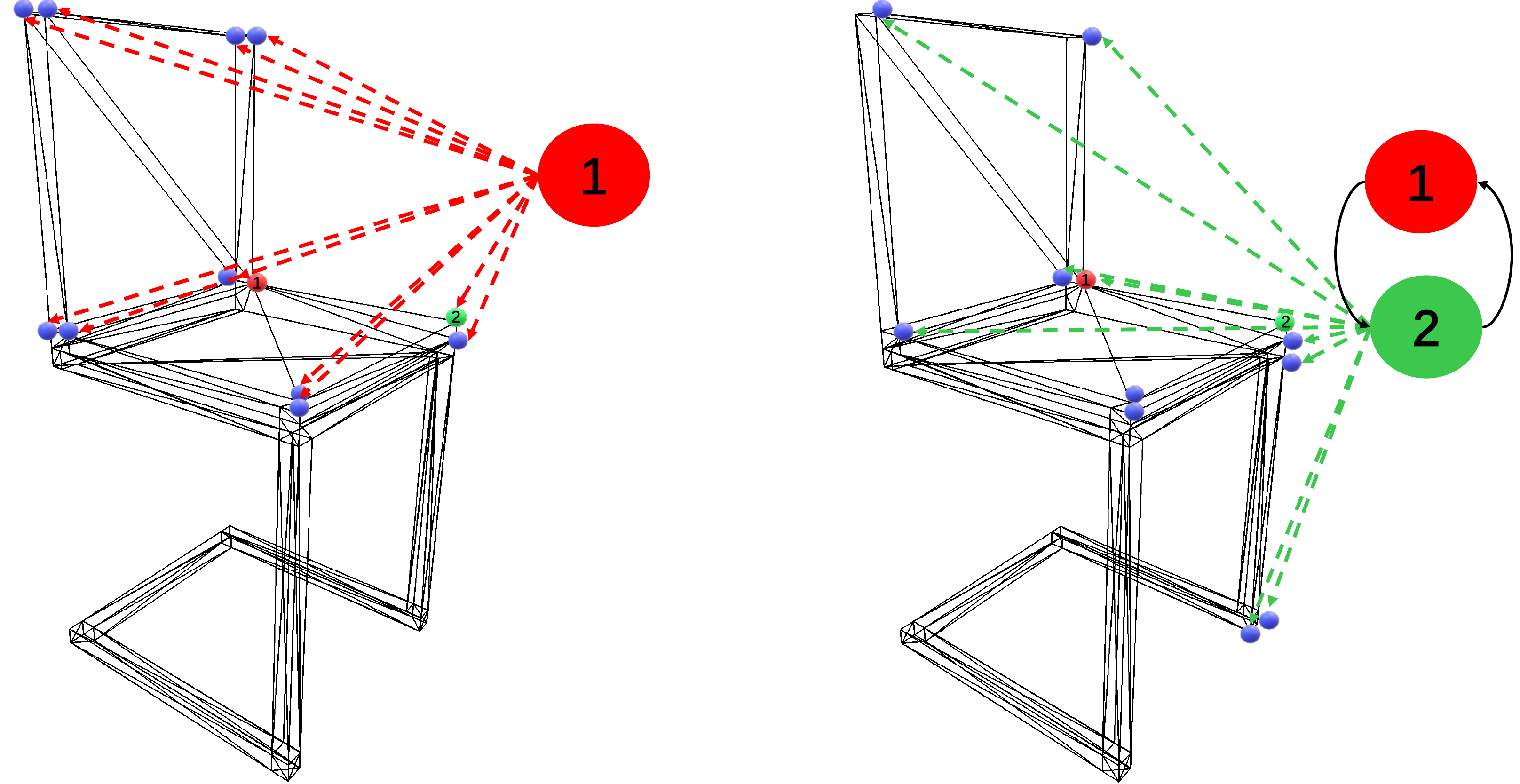}
  \end{center}
     \caption{The left (resp. right) figure shows that the convolution starts from vertex 1 (resp. 2). 
}
  \label{fig:tagcn}
\end{figure}
\begin{equation}
\mathcal{L}_{c d}=\sum_{x \in M} \min _{y \in S}\|x-y\|_{2}^{2}+\sum_{y \in S} \min _{x \in M}\|x-y\|_{2}^{2}
\end{equation}
where the $M, S$ are the two point sets.
Akin to \cite{wang:3DN,smith19a:GEOMetrics}, our method does not simply compute the CD loss between the predicted points and ground-truth points. Accordingly, we uniformly sample the same number of vertices from the predicted mesh and ground-truth mesh and then compute the CD loss between the predicted mesh and ground-truth mesh. More precisely (see Fig.\ref{fig:sample}), for each triangular face, we first compute its area and store it in an array along with the cumulative area of triangles visited so far. Next, we select a triangle with a probability ratio between its area and the total cumulative area. For each selected triangular face defined by vertices $v_1, v_2, v_3$, a new point $r$ can be sampled uniformly from the surface by the following formulation:
\begin{equation} 
    r = (1 - \sqrt{u})v_1 + \sqrt{u}(1 - w)v_2 + \sqrt{u}wv_3
\end{equation}
where $u, w \sim U(0, 1)$. 
Note that this formulation allows us to differentiate through random selection via the re-parametrization trick.

Since the CD loss does not concern the connectivity of mesh vertices, the predicted mesh could suffer from few floating vertices and self-intersections. 
Thus we add some geometric-aware regularizaton terms, such as a Laplacian loss and an edge length loss which prevent the vertices from moving with excessively long distance and can potentially avoid mesh self-intersection.
To calculate the Laplacian loss $l_1$, we first define a Laplacian coordinate 
for each vertex $p$ as $\delta_p = p - \sum_{k \in N(p)}\frac{1}{||N(p)||}k$, which is calculated as: $l_{lap} = \sum_p ||\delta^{'}_p - \delta_p||_2^2$, 
where $\delta^{'}_p$ is the Laplacian coordinate of a vertex $p$ after the deformation block, and $\delta_p$ is the input Laplacian coordinate of a vertex $p$.
The edge length loss is defined as $l_{edge} = \sum_p \sum_{q \in N(p)} || p - q||^2_2$ to make the predicted mesh visually appealing.

The total loss of deformation module is

\begin{equation}
  L_{all} = l_{cd} + \lambda_1l_{lap} + \lambda_2l_{edge}
  \label{Eq}
\end{equation}
where the $\lambda_1, \lambda_2$ are the hyper-parameters weighting the importance of each of them. Note that the loss $L_{all}$ is applied to the output of each mesh deformation block.

\section{Experiments}
In this section, we demonstrate the efficiency and performance of STD-Net on the single RGB images structure-preserving reconstruction by taking the benefits of the mesh-based structure representation. In addition, we present an ablation study to demonstrate how individual components of our model contribute to its overall performance. 

\subsection{Dataset}
We use ShapeNet \cite{shapenet2015} 
as the main dataset for training and testing the STD-Net. Each shape is rendered with 24 different views, and the rendered image resolution is $224 \times 224$. We use three categories for our experiment: Chair, Table and Airplane. Each shape is segmented according to their mesh components and then the bounding box of each part is generated. Each category was split into a training, validation, and test set with a ratio of 7:1:2 respectively, which is also similar to \cite{Wang2018:Pixel2Mesh} and \cite{smith19a:GEOMetrics}.

\subsection{ Implementation Details}

\subsubsection{Structure Recovery network}
For the encoder, the contour estimator consists of a VGG-16 (two fully connected layers) and one 9x9 convolution layer. The fusion phase is implemented by CNN. As for the decoder, we use the pre-trained shape abstraction decoder in StructureNet \cite{mo2019structurenet}. There are two stages in training the structure recovery network. We first train the network for estimating a binary object mask for the input image. The first and second scale network are trained jointly. Secondly, we train the encode and decoder together, during which a low learning rate for the encoder is used.

\subsubsection{Mesh Deformation Network}
Prior to train this network, we prepare the pairs of bounding box and mesh in ShapeNet. For each category, a bounding box(OBB)-to-mesh mapping is trained from the ground truth. These mappings are trained with Adam Optimizer ($\beta = 5e^{-4}$) and a learning rate of $3e^{-5}$. We conduct the training for $10^5$ iterations and empirically stop it, with the best model selected from evaluating on the validation set every 10 iterations. The hyper-parameters setting used, as described in Eq. \ref{Eq}, are $\lambda_1=0.3$, $\lambda_2=0.1$. 

\begin{figure}[]
  \begin{center}
     \includegraphics[width=1\linewidth]{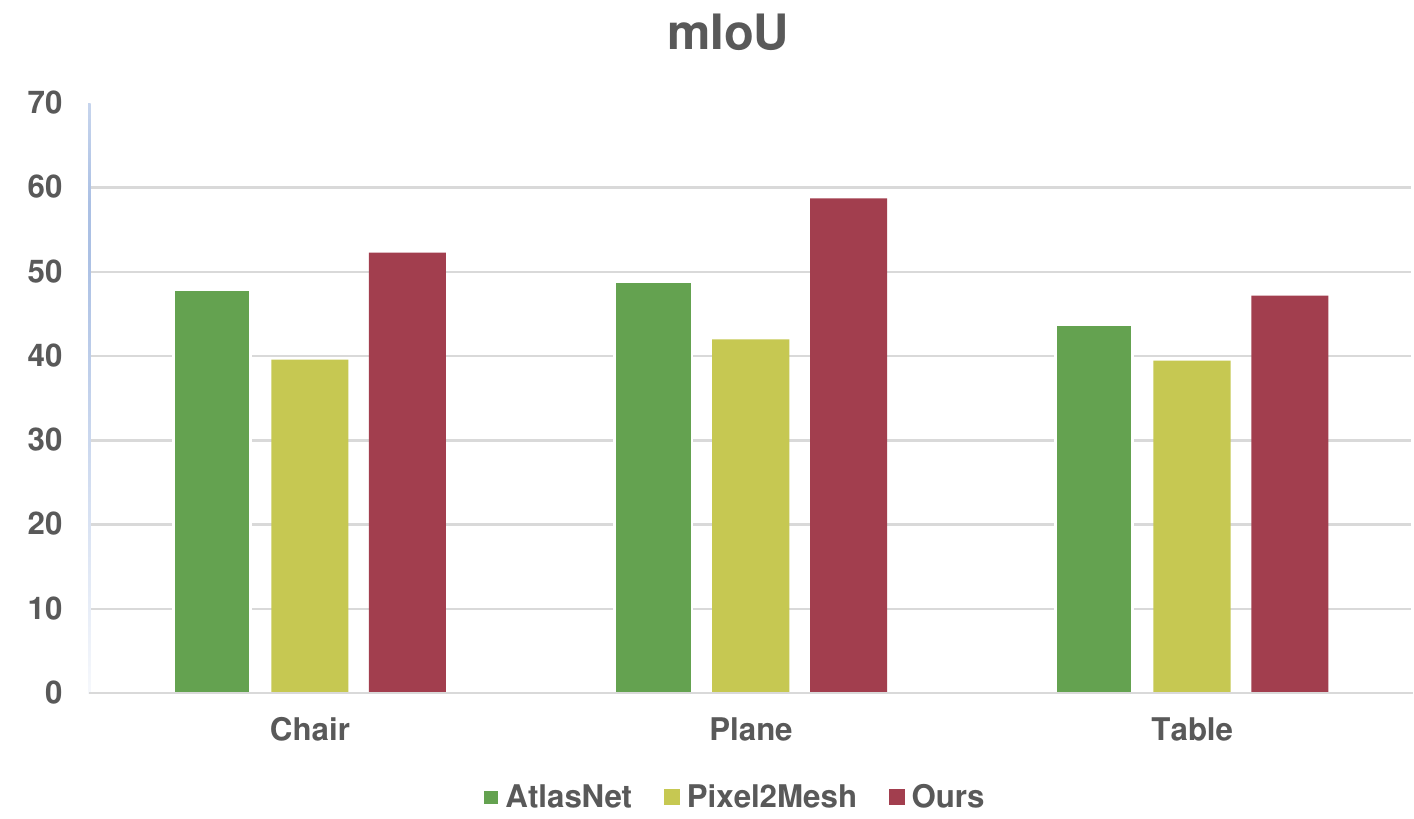}
  \end{center}
     \caption{Comparison of mean IoU($\%$) with other mesh-based approaches on the three categories of ShapeNet dataset.}
  \label{fig:mIoU}
  \end{figure}

\begin{table}
\begin{center}
\caption{The F1-score ($\%$) on the three categories at threshold $d = 1e-4$. Larger is better. Best results (our) are bold.}
\label{tab:f1-score}
\begin{tabular}{lllll}
\hline\noalign{\smallskip}
& Pixel2Mesh & AtlasNet & GEOMetric & ours\\
\noalign{\smallskip}
\hline
\noalign{\smallskip}
Chair & 54.38 &  55.2 & 56.61 & \bfseries{58.23}\\
Plane & 71.12 & 77.15 & 89.00   &  \bfseries{92.48}\\
Table & 66.30 & 59.49 & 66.33 & \bfseries{68.24} \\
mean & 63.9 & 63.61 & 70.64 & \bfseries{72.98}\\
\hline
\end{tabular}
\end{center}
\end{table}
\subsubsection{Comparison}
We evaluate the quantitative performance of our method by comparing with those results generated by the mesh-based approaches in Pixel2Mesh \cite{Wang2018:Pixel2Mesh},
AtlasNet\cite{Groueix2018:AtlasNet} and GEOMetrics \cite{smith19a:GEOMetrics}. 
To do this, we train our network on the ShapeNet and evaluate two metrics (F1 and mIoU) on test data sets with state-of-the-art approaches. 
We mainly test on the three categories in ShapeNet: Chair, Table and Airplane, and then perform the mainly common-used metrics on the shape reconstruction. First, we compare the F1 score, which is a harmonic mean of precision and recall at a given threshold of $d$. 
The results of comparison are summarized in Table \ref{tab:f1-score}. It shows that our method has much better performance than previous approaches on the three categories under a smaller threshold of $d=1e-4$. Second, we compare the mIoU with the other works, which is depicted in the Fig. \ref{fig:mIoU}, showing that our method get a high performance that others.

Qualitative comparison results with state-of-the-art works on shape reconstruction for each category are presented in Figure \ref{fig:result}. These results demonstrate that our method can provide highly accurate reconstruction of objects from the input RGB images and capture structure details effectively. 

\subsection{Ablation Study}
In this section, we study the influence of our network's components and demonstrate their importance by comparing the full method's results on the chair data with the ablated version of the network. Here, we mainly evaluate the impact of the topological-adaptive layers by replacing them with Naive GCN. The qualitative results are shown in Table~\ref{table:ablation}. 

\begin{table}
\begin{center}
\caption{An ablation study of mesh deformation network.}
\label{table:ablation}
\begin{tabular}{llll}
\hline\noalign{\smallskip}
 & Chamfer loss & F1-score & IoU\\
\noalign{\smallskip}
\hline
\noalign{\smallskip}
Naive GCN & 3.550 & 30.84 & 12.76  \\
TAGCN & \bfseries{1.390} &  \bfseries{72.98} & \bfseries{52.6}\\
\hline
\end{tabular}
\end{center}
\end{table}

\section{Conclusions}
In this paper, we propose a structure-preserving and topological-adaptive deformation network for 3D objects reconstruction from single images. Our method provides a graph representation of 3D models with cuboid bounding boxes, which can delicately describe the structure information of the object, and thus can reconstruct 3D models with complex structure directly from the single image. Our proposed learning framework consists of a structure recovery network and a mesh deformation network. The former is designed with an auto-encoder which generates cuboid bounding boxes for the object, and the latter consists of three deformation blocks intersected by two graph unpooling layers, which progressively deform the input meshed bounding box into the meshed models. The most significant feature of the mesh deformation network is that it accepts the input graph with different topology types, which enables the reconstruction of 3D models with various complex structures. Compared to previous works, our method could achieve better performance in 3D shape structure-preserving  reconstruction.

\setlength{\tabcolsep}{4pt}

\setlength{\tabcolsep}{1.4pt}

\clearpage
%
%
\bibliographystyle{splncs04}
\bibliography{eccv2020submission}
\end{document}